\documentclass[twocolumn]{aastex62}
\usepackage{graphicx}
\usepackage{amsmath}
\usepackage{comment}
\usepackage{cases}
\usepackage{color}

\DeclareMathAlphabet{\mathsc}{OT1}{cmr}{m}{sc}
\def\testbx{bx}%
\DeclareRobustCommand{\ion}[2]{%
\relax\ifmmode
\ifx\testbx\f@series
{\mathbf{#1\,\mathsc{#2}}}\else
{\mathrm{#1\,\mathsc{#2}}}\fi
\else\textup{#1\,{\mdseries\textsc{#2}}}%
\fi}

\newcommand{\beq}{\begin{equation}}
\newcommand{\eeq}{\end{equation}}

\newcommand{\hi}{H{\sc i}~}
\newcommand{\HI}{H{\sc i}}

\newcommand{\NaI}{Na~{\sc i}~}

\newcommand{\kms}{km ${\rm s^{-1}}$~}

\newcommand{\nhi}{$\mathrm{N}_\mathrm{H{\sc I}}$}
\newcommand{\nHI}{$\mathrm{N}_\mathrm{H{\sc I}}~$}

\begin{document}

\title{Small-Scale HI Channel Map Structure is Cold: Evidence from \NaI Absorption at High Galactic Latitudes}

\author[0000-0003-4797-7030]{J.E.G. Peek}
\affiliation{Space Telescope Science Institute, 3700 San Martin Dr, Baltimore, MD 21218, USA}
\affiliation{Department of Physics \& Astronomy, Johns Hopkins University, Baltimore, MD 21218, USA}
\author[0000-0002-7633-3376]{S.E. Clark}
\affiliation{Institute for Advanced Study, 1 Einstein Drive, Princeton, NJ 08540, USA}
\altaffiliation{Hubble Fellow}
\email{seclark@ias.edu}

\begin{abstract}
The spatial distribution of neutral hydrogen (\HI) emission is a powerful probe of interstellar medium physics. The small-scale structure in \hi channel maps is often assumed to probe the velocity field rather than real density structures. In this work we directly test this assumption, using high-resolution GALFA-\hi observations and 50,985 quasar spectra from SDSS. We measure the equivalent widths of interstellar \NaI D$_1$ and \NaI D$_2$ absorption, and robustly conclude that they depend more strongly on the column density of small-scale structure in \hi than on either the large-scale \hi structure or the total \hi column. This is inconsistent with the hypothesis that small-scale channel map structure is driven by velocity crowding. Instead, the data favor the interpretation that this emission structure predominantly originates in cold, dense interstellar material, consistent with a clumpy cold neutral medium.

\end{abstract}

\section{Introduction}

The diffuse interstellar medium (ISM) is sculpted by the interplay of many thermodynamic, magnetohydrodynamic, and chemical processes, and contains structure over a large range of scales. The morphology of diffuse gas and dust is an important diagnostic of the physics of the ISM. Small-scale structure in neutral hydrogen (\HI) emission is of particular interest because it is preferentially aligned with the magnetic field, making the structure of \hi a unique probe of the interstellar magnetic field \citep{Clark:2014, Clark:2015, Martin:2015tr, Kalberla:2016, Blagrave:2017, Clark:2018}. Filamentary, magnetically aligned structures are also observed in other ISM probes, including \hi absorption \citep{McClureGriffiths:2006wx}, dust emission \citep{PlanckXXXII}, and diffuse synchrotron emission \citep{Jelic:2015}.

Interstellar \hi is generically distributed into three phases, a warm neutral medium (WNM), cold neutral medium (CNM), and a thermally unstable medium \citep[e.g.][]{Cox:2005, kalberla09}. While the linear \hi self-absorption structures discovered by \citet{McClureGriffiths:2006wx} are unambiguously CNM, with spin temperatures of $\sim 40 \,{\rm K}$, all three thermal phases can contribute to the observed \hi emission. The magnetically aligned \hi structures are most prominent in narrow channel maps: \hi emission integrated over a few \kms in line velocity. The data thus far favor the interpretation that the magnetically aligned \hi emission structures are preferentially CNM.
Linewidth measurements of the \hi fibers are consistent with CNM temperatures \citep{Clark:2014, Kalberla:2016}. 

\citet{Clark:2019} went further, showing that the small-scale structure in \hi channel maps \textit{in general} -- not just the prominently linear structure -- is preferentially associated with cold-phase material. 
That work tested and ruled out a pervasive claim in the literature: that small-scale structure in \hi channel map emission is dominantly or entirely ``caustics" imprinted by the turbulent velocity field. The notion that \hi channel map structure traces the velocity field leads to the incorrect conclusion that the magnetically aligned \hi structures are neither CNM nor real density structures at all. This belief is summarized in \citet{LazarianYuen:2018}: ``On the basis of \citet{LP00}, one can conclude 
that the filaments observed by \citet{Clark:2015} in thin channel maps can be identified with caustics caused by velocity crowding." Indeed, those authors' intepretation of \citet{LP00} is that in thin velocity channel maps ``most of the structures should be due to velocity caustics". Instead, \citet{Clark:2019} showed that 1. the \hi channel map structures are correlated with broadband FIR emission, which is not sensitive to the velocity field, 2. that this correlation does not measurably decrease as the velocity channel width is decreased, and 3. that the FIR/\nHI ratio is preferentially higher toward smaller-scale channel map structures. None of these findings are consistent with measurable velocity caustics in \HI. 

Additionally, \citet{Clark:2019} challenged the idea that the spatial power spectrum of narrow \hi channel maps can be used to measure the spectral index of the velocity power spectrum  also \cite[see also][]{Kalberla:2019de}. If narrow channel maps were structured by the velocity field, a shallower power spectrum would be measured for narrower channels \citep{LP00}. The converse of this argument holds that because narrower \hi channels show shallower power spectra, the thin channel maps must be patterned by the velocity field. However, as \citet{Clark:2019} points out, shallower power spectra in narrower channels is also qualitatively consistent with a higher fraction of the emission in narrow channel maps originating in CNM structures, which are smaller-scale than the more diffuse WNM.

In this work, we carry out an independent test of the physical nature of this structure, using interstellar absorption lines in SDSS quasar spectra. Specifically, we measure the behavior of the equivalent width (EW) of the \NaI absorption line as a function of how much small-scale HI channel map \nHI the quasar sightlines traverse. \NaI absorption is a direct tracer of cooler material in the ISM. The ratio $n_{\sc Na I}/n_{\sc H I}$ will increase in denser regions due to the $n_e\times n_{\rm ion}$ scaling of recombination and in cooler regions due to the $T^{-0.7}$ scaling of the radiative recombination  coefficient \citep{1975ApJ...202..628H,2011piim.book.....D,1978ppim.book.....S}. If small-scale HI channel map structure is preferentially correlated with cold, dense regions of gas, then controlling for total \hi column, the column of \NaI and thus its EW will be higher for quasar sightlines that pass through more small-scale structure. If instead the small-scale HI channel map structure is ``caustics caused by velocity crowding" \citep{LazarianYuen:2018}, we would expect no increase in EW.

This Letter is organized as follows. In Section \ref{s:data} we introduce the data products used. In Section \ref{s:analysis} we detail our analysis of the correlation between small-scale \hi channel map structure and \NaI absorption measurements. In Section \ref{s:discussion} we discuss how these results fit into a broader picture of ISM structure, and in Section \ref{s:conclusion} we conclude with implications of this work for statistical diagnostics of the turbulent ISM. 

\begin{figure*}
\begin{center}
\includegraphics[scale=1.0]{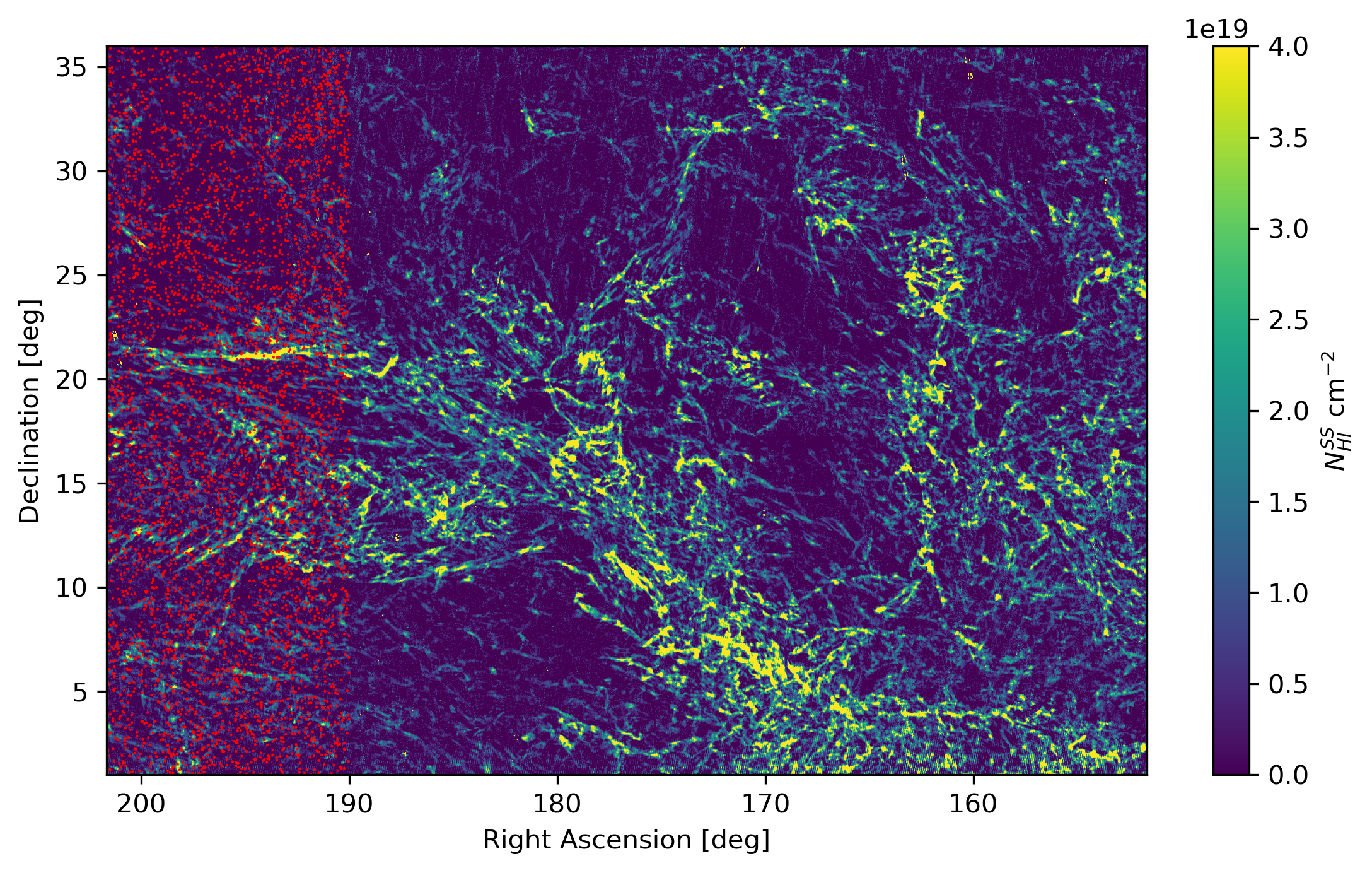}
\caption{A section of the $N_{\rm HI}^{SS}$ map at high Galactic latitude. The locations of quasars in our sample are shown in red for $l>190^\circ$, to give a sense of the density of the quasar sample on the sky without obscuring the rest of the map.}
\label{f:NHI_SS}
\end{center}
\end{figure*}

\section{Data}\label{s:data}
\subsection{GALFA-HI}
We use archival 21-cm observations of the Milky Way ISM from GALFA-HI \citep{2018ApJS..234....2P}. These data, obtained with the 305m Arecibo telescope in Puerto Rico, represent the highest angular ($4^\prime$) and spectral ($0.18$ km s$^{-1}$) resolution large area (4 steradians, -1$^{\circ}~<\delta<~38^\circ$) survey of the Galactic ISM to date. The data have been corrected for small angular scale sidelobe effects and spectral fixed pattern noise. We use the main  $\delta v$ = 0.736 km s$^{-1}$ (Wide) data cubes as well as the total integrated \hi intensity from -90 km s$^{-1}~ <~v_{\rm LSR}~<$ 90 km s$^{-1}$, which has been corrected for stray radiation. We note that we use the data cubes for small scale spatial information, so the lack of stray correction for these data, which entirely effects larger angular scales, is not a concern.

\subsection{SDSS}
We use archival spectroscopic observations of quasars from Sloan Digital Sky Survey Data Release 7 \citep[SDSS DR7;][]{Abazajian:2009ef}. These data have been used many times for spectroscopic absorption line studies of diffuse material, including of \NaI D$_1$ \& D$_2$ absorption in the Milky Way \citep{Murga:2015bh,Poznanski:2012bf}. Here we take advantage of the published continua as well, derived using non-negative matrix factorization in \cite{Zhu:2014gf}. We access these data using the igmspec database published in \cite{Prochaska:2017et}. The entire database contains 105,783 quasars, with a wavelength resolution of $\lambda/\delta \lambda$ = 1800.

\section{Analysis}\label{s:analysis}

\subsection{HI structure}\label{s:hi}
We construct a map of the small angular scale structure in the 21-cm data from GALFA-HI. To do this we first construct 27 channel maps of the 21-cm data with 3.68 km s$^{-1}$ velocity width that span the velocity range $-48.21 ~ {\rm km~s}^{-1}< v_{\rm LSR} <  50.42 ~ \rm{km~s}^{-1}$. These maps cover the whole sky available in the data set. This velocity range includes nearly all Galactic emission and all significant small-scale structures at these latitudes; velocities beyond this range tend only to include noise and artifacts. We convert these maps to column densities under the assumption of optically thin \HI, which is largely a good assumption at high Galactic latitude \citep{Murray:2018}, where 
\begin{equation}
N_{\rm HI} = 1.8 \times 10^{18}~ {\rm cm}^{-2} \frac{\int T_B dv}{\rm K~ km~s^{-1}}.
\end{equation}
We then perform an unsharp mask on each of these maps. This procedure consists of smoothing the maps with a Gaussian of width FWHM=$30'$, and subtracting the smooth map from the original data. It is equivalent to a Gaussian high-pass Fourier filter. We then threshold each of these maps, setting all values below 2$\times 10^{18}{\rm ~cm}^{-2}$ to zero, as a way to suppress artifacts caused by the observational scan pattern. We truncate the maps to the range 1$^{\circ}~<\delta<~36^\circ$ to reject some artifacts generated by the Arecibo ground screen \citep{Peek:2011fp}, and we restrict the area to Galactic latitude $|b|>30^\circ$. We refer to this small-scale structure column density map as $N_{\rm HI}^{SS}$ (see Figure \ref{f:NHI_SS}). We perform the same $\delta$ and $b$ cuts on the stray-corrected column density map, $N_{\rm HI}$, and define a large-scale structure column density map as
\begin{equation}\label{e:fit}
N_{\rm HI}^{LS} =  N_{\rm HI}-N_{\rm HI}^{SS}.
\end{equation}

$N_{\rm HI}^{LS}$ is ``large-scale" in the sense that it is all of the \hi column that is not in $<30'$ structures in the channel maps.

\subsection{Quasar Spectra}
We take the complete list of 105,783 quasars from SDSS DR7 and restrict to sources that are in the footprint described in \S \ref{s:hi}, which gives us a final list of 50,985 quasar spectra. We interpolate these spectra, along with the paired continua from \citep{Zhu:2014gf}, onto a common wavelength grid from 5860\AA~to 5940\AA~in steps of 0.4\AA.

\begin{figure*}
\begin{center}
\includegraphics[scale=0.85]{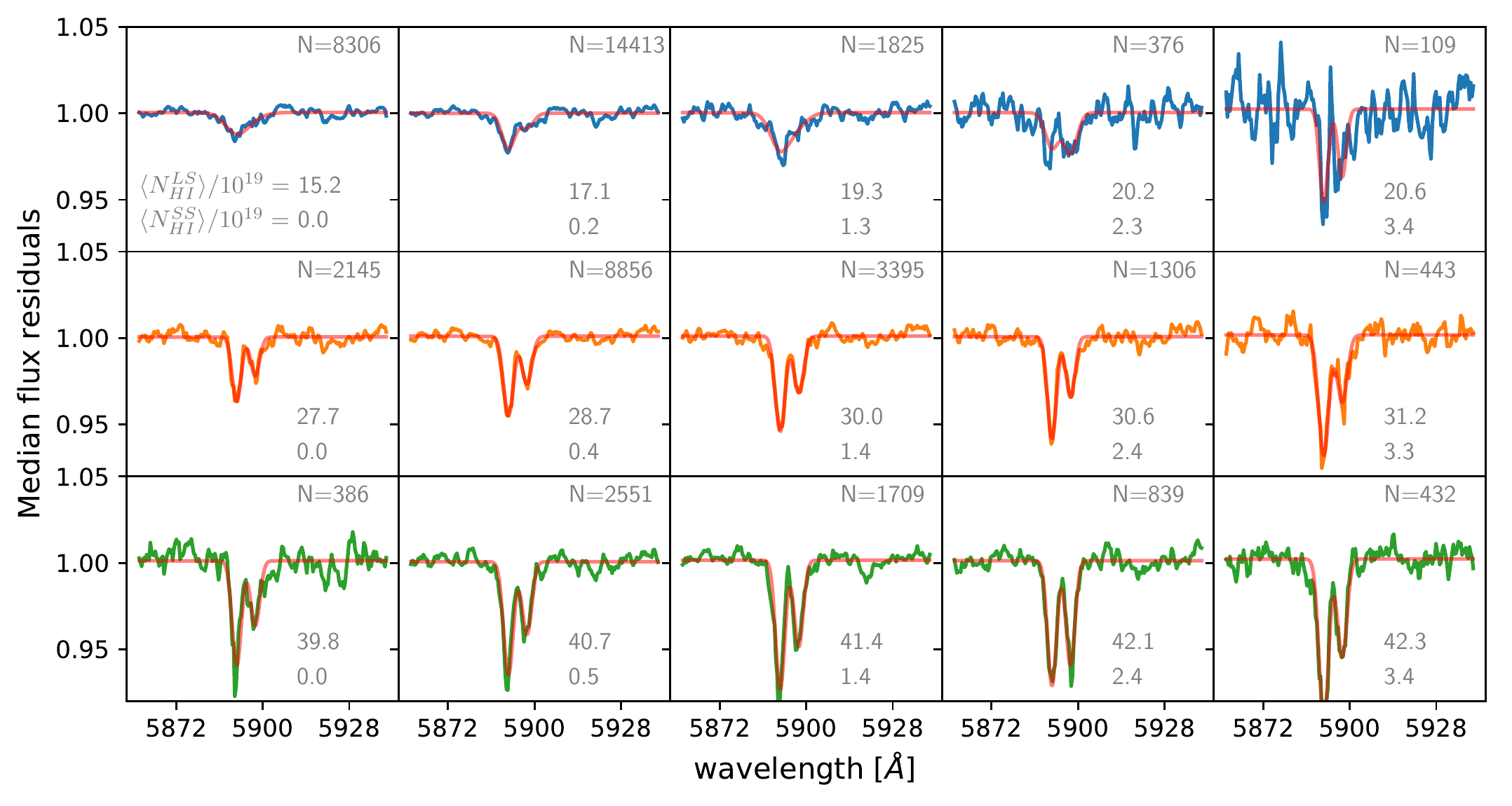}
\caption{Median stacks of \NaI $D_1$ and $D_2$ absorption lines. Each panel is a median stack (blue, orange, or green) with a 2-Gaussian fit overlaid (red). The leftmost column represents stacks where $N_{\rm HI}^{SS}$ = 0. The second through fifth columns represent stacks observed along lines of sight with N$_{HI,i}<$ N$_{HI, U} \le $N$_{HI,i+1}$, where N$_{HI,i}$ = [0, 1, 2, 3, 4]$\times 10^{19}$ cm$^{-2}$. The $j$th row stacks spectra observed along lines of sight with N$_{HI,j}<$ N$_{HI, S} \le$N$_{HI,j+1}$, where N$_{HI,j}$ = [1, 2.33, 3.66, 5]$\times 10^{20}$ cm$^{-2}$. The number of medianed spectra is marked in the upper right of each box, and the values of the median of the column densities ($\langle N_{HI}^{LS}\rangle$ and $\langle N_{HI}^{SS}\rangle$) are shown in the lower right.}
\label{f:grid_meds}
\end{center}
\end{figure*}

\begin{figure*}
\begin{center}
\includegraphics[scale=0.9]{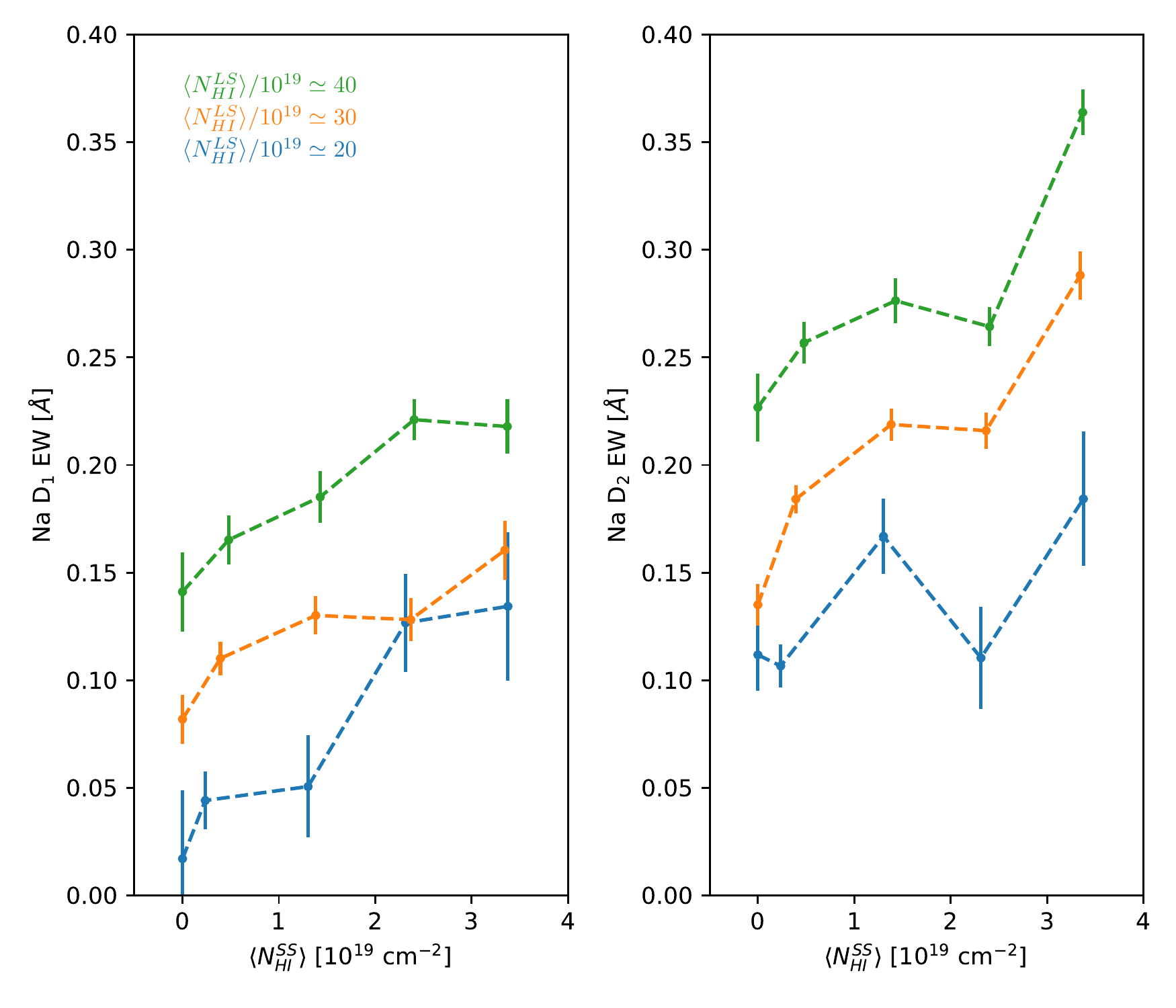}
\caption{Fits to the equivalent width of the median stacks shown in Figure \ref{f:grid_meds}. Blue, orange, and green points represent the low, medium, and high $\langle N_{HI}^{LS} \rangle$ described in \S \ref{s:analysis}. Bins are located on the abscissa by their median small-scale column, $\langle N_{HI}^{SS}\rangle$.}
\label{f:EW_2fits}
\end{center}
\end{figure*}

\subsection{Stacking}

Following \cite{Poznanski:2012bf} and \cite{Murga:2015bh}, we perform median stacking of the continuum normalized, interpolated spectra. Median stacking has the strong advantage of rejecting outliers and the effects of incorrect telluric calibrations; indeed, we find mean stacking creates significantly noisier spectra. We find our results are qualitatively insensitive to choosing means or medians (see \S \ref{s:discussion}). We stack the quasar spectra in bins of the $N_{\rm HI}^{SS}$ and $N_{\rm HI}^{LS}$. We bin the data against $N_{\rm HI}^{LS}$ in three bins (low, medium, and high), with the 4 bin edges equally spaced from 10$^{20}$ cm$^{-2}$ to 5 $\times$ 10$^{20}$ cm$^{-2}$. We bin the data against $N_{\rm HI}^{SS}$ in 5 bins. Four of these bins have bin edges evenly spaced from zero to 4 $\times$ 10$^{19}$ cm$^{-2}$. We also include a fifth bin for $N_{\rm HI}^{SS}$ = 0, which has the largest number of sources. This creates a total of 3$\times$5 = 15 bins. We select this binning to have sufficient \NaI D$_1$ and \NaI D$_2$ signal in each bin to make a reasonable measurement, while also including most (92\%) of the sources. We note that perturbations to this binning do not qualitatively impact our results.

\subsection{Fitting}\label{fitting}

We fit the \NaI D$_1$ and \NaI D$_2$ absorption lines in each of these median-binned spectra (Figure \ref{f:grid_meds}). We fit two Gaussians with independent amplitudes, but with wavelengths fixed to the rest frequency and only one width parameter used for both \NaI D$_1$ and \NaI D$_2$. 
We find the residual continua are generally quite well behaved in the medians, and fit only an additional overall offset parameter. We use the covariance between fit parameters to estimate the errors on the equivalent widths on each line and the system of both lines. Equivalent width fits to \NaI D$_1$ and \NaI D$_2$ are shown in Figure \ref{f:EW_2fits}. In bins where both lines are robustly and clearly detected, the line ratio is between 1.6 and 1.35, indicating some saturation in the \NaI D$_2$ line.

\begin{figure}
\begin{center}
\includegraphics[scale=0.8]{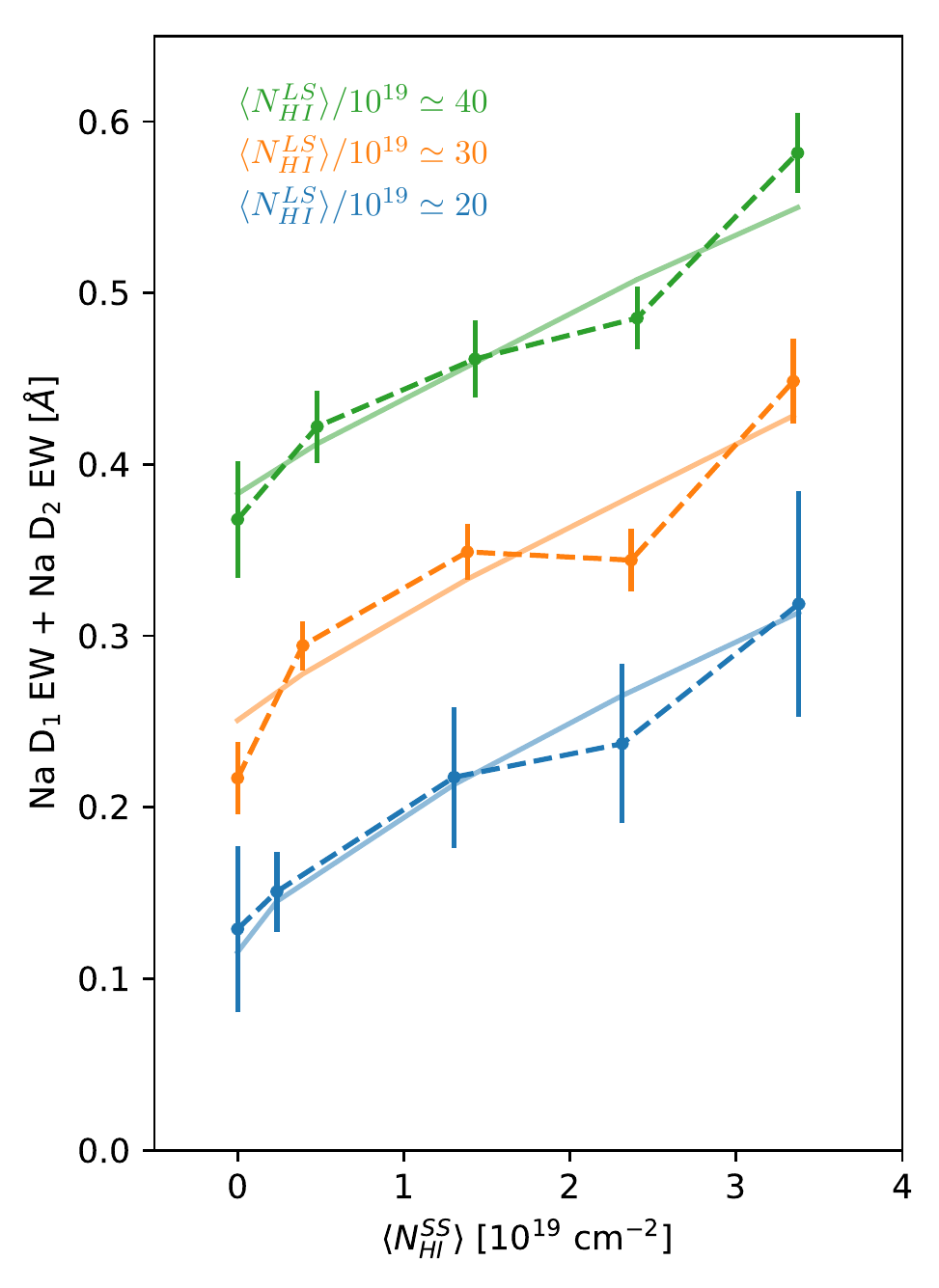}
\caption{The integrated \NaI D equivalent widths from Figure \ref{f:EW_fits}. Fits to Equation \ref{e:fit} are overplotted in solid lines. Bins are located on the abscissa by their median small-scale column, $\langle N_{HI}^{SS}\rangle$.}
\label{f:EW_fits}
\end{center}
\end{figure}

It is evident that in the noisiest bins with the fewest sources the negative covariance between \NaI D$_1$ and \NaI D$_2$ can generate significant noise. We therefore also compile a measurement of the total equivalent width of \NaI D$_1$ and \NaI D$_2$, shown in Figure \ref{f:EW_fits}. These data are then fit with a simple model,

\begin{align}\label{e:fit}
\langle{\rm EW}_{\rm Na~ I~ D_1}\rangle + \langle{\rm EW}_{\rm Na~ I~ D_2}\rangle  &=  C^{LS}\frac{\langle N_{HI}^{LS}\rangle}{10^{19}~ \rm cm^{-2}}  \nonumber \\ 
&+  C^{SS}\frac{\langle N_{HI}^{SS}\rangle}{10^{19}~ \rm cm^{-2}} + EW_{0},
\end{align}

as shown in Figure \ref{f:EW_fits}. We find $C^{SS} = 41.5 \pm 6.1$ m\AA, and $C^{LS} = 10.8 \pm 0.8~$ m\AA~ and $EW_{0} = -50 \pm 27$ m\AA.

\subsection{Variations}

This analysis makes many specific choices, but the basic result, that $C^{SS}$ is significantly larger than $C^{LS}$, is quite robust. The total EW of \NaI is much more sensitive to the column density of small scale structure than of large scale structure. We find this same result if we switch to using means instead of medians, if we vary the bin edges and numbers, and if we use the total $N_{HI}$ as a proxy for $N_{HI}^{LS}$. We can even conduct the analysis without binning at all, and simply integrate over the \NaI D$_1$ and \NaI D$_2$ lines to find the total equivalent width for each line of sight: we still find the same qualitative result. The procedure described in previous sections was selected because it allows for the easiest visual confirmation of our results, and follows previous work with these data \citep{Poznanski:2012bf, Murga:2015bh}.

 The Rolling Hough Transform \citep[RHT;][]{Clark:2014} was designed to find linear features, whose properties in these same data were shown to be consistent with CNM. The RHT is a sequence of imaging processing steps, with the USM as the first step. Thus, it is not at all surprising that over the region we examine in this work 87\% of the RHT-detected pixels overlap with the USM pixels described in \ref{s:hi}. (RHT-detected pixels are simply the pixels with nonzero RHT intensity as described in \cite{Clark:2014} for the data in \ref{s:hi}, plus weak thresholding to suppress imaging artifacts.) Indeed, we see the expected trend when we examine pixels highlighted by the RHT: controlling for total \hi column, these magnetically aligned structures are higher in \NaI D EW. 

\section{Discussion}\label{s:discussion}

This analysis shows that the equivalent width of interstellar \NaI D$_1$ and \NaI D$_2$ absorption is significantly more strongly dependent on the column density of small-scale structure in neutral hydrogen, $N_{HI}^{SS}$, than it is on the column density of larger-scale structure $N_{HI}^{LS}$ or the overall column $N_{HI}$. This is quite consistent with a picture in which this small scale structure represents the cooler, denser part of the ISM, as we have suggested previously \citep{Clark:2014, Clark:2019}. \NaI / \nhi~ will increase in denser regions due to the $n^2$ scaling of recombination. \NaI / \nhi~ will also increase in cooler regions due to the $T^{-0.7}$ scaling of the radiative recombination coefficient \citep{2011piim.book.....D}. This leads to a higher \NaI column (as probed with the \NaI D$_1$ and \NaI D$_2$ lines) per unit $N_{HI}$ than the typical ISM, which produces the trend $C^{SS} > C^{LS}$.

In contrast, a velocity crowding origin for the small \hi structure would yield the opposite trend. If the state of the ISM is not related to the intensity of smaller structures in the neutral ISM, the trends between \NaI D$_1$ and \NaI D$_2$ EW and \nHI should generically be independent of whether the \nHI is in smaller structures or larger structures. Further, if small-scale structure is created by velocity crowding, \NaI D absorbers should likewise be crowded into similar velocities. This will naturally enhance the optical depth of the stronger saturated line, which will in turn drive down the measured EW per \nhi. Thus a velocity crowding origin predicts $C^{SS} < C^{LS}$, which is inconsistent with our result.

\section{Conclusion}\label{s:conclusion}

We stacked 50,985 SDSS DR7 spectra at high Galactic latitude, binned against the smooth \hi column density $N_{HI}^{LS}$ and small scale structure column density in thin velocity channels, $N_{HI}^{SS}$, as measured by GALFA-\hi DR2. We find the \NaI D$_1$ and \NaI D$_2$ EW measured in these median stacks have a much stronger dependence on $N_{HI}^{SS}$ than on $N_{HI}^{LS}$. This is consistent with a picture in which small \hi structure is dominated by cool, dense media, but inconsistent with a picture in which the small structure is velocity crowding.

There are two main upshots of this result. The first is conclusive proof that small structures in velocity slices of 21-cm measurements indeed have a significant overabundance of cooler, denser ISM. The exact composition of this material is still unknown, and we have not measured the relative balance of warm, unstable, cold, diffuse molecular, and dense molecular gas. A more precise measurement with higher spectral resolution and using a broader range of ions would be very illuminating in terms of the precise nature of the gas in question, perhaps using the KODIAQ archive \citep{2017AJ....154..114O}, the Hubble Spectroscopic Legacy Archive \citep{2017cos..rept....4P}, and COS-GAL \citep{2019ApJ...871...35Z}. 

The second upshot is that a number of statistical tools developed for the study of ISM turbulence and magnetization must be dramatically rethought. Many methods, e.g. Velocity Channel Analysis \citep[VCA;][]{LP00}, rely on the notion that small scale structures are driven by velocity crowding effects. While it is certain that the details of which structures appear at which velocities will be affected by the divergence and convergence of the velocity field along the line of sight, it is clearly unacceptable to use statistical tools that entirely ignore the fact that these structures are preferentially cool, dense media. The high wavenumber end of the narrow-channel \hi spatial power spectrum probes the density distribution of cold gas. The relative prominence of cold gas in narrower velocity channels will flatten the slope of the spatial power spectrum, creating the effect some have attributed to the velocity field. 
Thus, tools like VCA will need to be overhauled before they can be reliably used to measure physical properties of the ISM.

\software{astropy \citep{Astropy:2013, Astropy:2018}, matplotlib \citep{Matplotlib:2007}, numpy \citep{Oliphant:2015:GN:2886196}}

\acknowledgements

We thank Ed Jenkins for very useful discussions on the nature of \NaI absorbers and Marc-Antoine Miville-Desch\^enes for thoughtful comments.

S.E.C. is supported by NASA through Hubble Fellowship grant \#HST-HF2-51389.001-A awarded by the Space Telescope Science Institute, which is operated by the Association of Universities for Research in Astronomy, Inc., for NASA, under contract NAS5-26555. 

This publication utilizes data from Galactic ALFA \hi (GALFA-\HI) survey data set obtained with the Arecibo L-band Feed Array (ALFA) on the Arecibo 305m telescope. The Arecibo Observatory is operated by SRI International under a cooperative agreement with the National Science Foundation (AST-1100968), and in alliance with Ana G. M\'endez-Universidad Metropolitana, and the Universities Space Research Association. The GALFA-\hi surveys have been funded by the NSF through grants to Columbia University, the University of Wisconsin, and the University of California.

\end{document}